\documentclass[a4paper,10pt]{article}
\usepackage[utf8x]{inputenc}
\usepackage{graphicx}
\usepackage{setspace}
\usepackage{a4,amsmath,amsfonts,latexsym,amssymb,amsthm}
\usepackage{hyperref}
\usepackage{tikz}
\usepackage{mathptmx}
\usepackage{amsmath,amscd}
\usepackage{multicol}

\usepackage{pgfplots}

\usepgflibrary{shapes.geometric} 
\usepgflibrary[shapes.geometric] 
\usetikzlibrary{shapes.geometric} 
\usetikzlibrary[shapes.geometric] 

\newcommand{\diag}{\mathop{\mathrm{diag}}}

\title{Novel Kac-Moody-type affine extensions of non-crystallographic Coxeter groups}
\author{Pierre-Philippe Dechant$^{1,2}$ \footnote{\noindent E-mail: {\tt pierre-philippe.dechant@durham.ac.uk}} , C\'eline B\oe hm$^2$ \footnote{\noindent E-mail: {\tt c.m.boehm@durham.ac.uk}} \  and   Reidun Twarock$^1$ \footnote{\noindent E-mail: {\tt rt507@york.ac.uk}} \\
$^1$ \emph{Department of Mathematics, York Centre for Complex Systems Analysis,} \\\emph{University of York, Heslington, York, UK,}\\ 
$^2$ \emph{Ogden Centre for Fundamental Physics, Department of Physics,} \\\emph{University of Durham, South Rd, Durham DH1 3LE, UK}\\}

\begin{document}

\maketitle

\begin{abstract}
 {Motivated by recent results in mathematical virology, we present} novel asymmetric $\mathbb{Z}[\tau]$-integer-valued affine extensions of the non-crystallographic Coxeter groups $H_2$, $H_3$ and $H_4$ derived in a Kac-Moody-type formalism. {In particular, we show that the  affine reflection planes which extend the Coxeter group $H_3$ generate (twist) translations along 2-, 3- and 5-fold axes of icosahedral symmetry, and we classify these translations in terms of the Fibonacci recursion relation applied to different start values. We thus provide an explanation of previous results concerning affine extensions of icosahedral symmetry in a Coxeter group context, and   extend this analysis to the case of the non-crystallographic Coxeter groups $H_2$ and $H_4$. These results will enable new applications of group theory in physics (quasicrystals), biology (viruses) and chemistry (fullerenes).} 


IPPP/11/66, DCPT/11/132 \\                                                                                                     
\end{abstract}

\section{Introduction}\label{intro}

Coxeter groups are abstract groups describable in terms of mirror symmetries \cite{Coxeter1934discretegroups}. The finite Coxeter groups correspond to the finite Euclidean reflection groups and were classified in Ref.~\cite{Coxeter1935Enumeration}. They include the symmetry groups of the Platonic solids as well as the Weyl groups of simple Lie algebras. A subset of these groups are non-crystallographic, i.e. they describe symmetries that are not compatible with lattices in two, three and four dimensions. The latter include the groups $H_2$, $H_3$ and $H_4$, which are the only Coxeter groups generating rotational symmetries of order 5. We are particularly interested in $H_3$ because it contains as a subgroup the rotational symmetries of the icosahedral group $I$, which is {of crucial importance for} the modelling of viruses in biology \cite{Stockley2010emerging, Caspar:1962, Twarock:2006b, Janner:2006b, Zandi:2004}, fullerenes in chemistry \cite{Kroto:1985,Kroto:1992, Twarock:2002b, Kustov:2008} and {quasicrystals in physics \cite{Katz:1989, Senechal:1996, MoodyPatera:1993b, Levitov:1988, Shechtman:1984}.} 
Finite Coxeter groups describe the  properties of these structures, e.g. of a viral protein container or a carbon onion, at a given radial level. In order to obtain information on how  structural properties at different radial levels are collectively constrained by symmetry, affine extensions of these groups need to be considered. Such affine extended, infinite Coxeter groups can be constructed from the finite ones by introducing affine reflections, i.e. reflection planes not passing through the origin. Examples of {such affine extensions of finite}{} Coxeter groups are the Weyl groups of infinite-dimensional Kac-Moody algebras. While infinite {counterparts} to the crystallographic finite Coxeter groups have been intensively studied \cite{Humphreys1990Coxeter}, much less is known for the non-crystallographic {equivalents}.   

In Ref.~\cite{Twarock:2002a} affine extensions of the non-crystallographic Coxeter groups $H_2$, $H_3$ and $H_4$ have been constructed based on an extension of their Cartan matrices following the Kac-Moody approach in Lie Theory. Under the assumption that the extended matrix is symmetric, a unique affine extension has been obtained in each case. In Ref.~\cite{Keef:2009} it was shown  that this approach is {not sufficient for applications in virology}: The structures of viruses follow several different extensions of $I$ by translation operators.
Motivated by these results, we revisit here affine extensions of the non-crystallographic Coxeter groups $H_2$, $H_3$ and $H_4$, and present a more general framework that accommodates the results of Ref.~\cite{Keef:2009}. In particular, we  relax the assumption that the extended Cartan matrix must be symmetric {and show that this leads to new affine extensions of $H_3$}. For those extensions that encode affine reflections at planes parallel to those of $H_3$, the affine reflections introduced in this way 
generate translations of various lengths when combined with the reflections encoded by $H_3$; in addition, there are other extensions that generate screw translations along the three- and five-fold axes of icosahedral symmetry. These results open up a new view on infinite-dimensional analogues to non-crystallographic Coxeter groups. 

The paper is organised as follows: {In Section \ref{sec_prelim}, we review the procedure of affine extension of crystallographic and 
non-crystallographic Coxeter groups based on Cartan matrices. In Section \ref{sec_new}, we construct novel affine extensions of non-crystallographic Coxeter groups by relaxing the symmetry condition on the Cartan matrix, and give a concrete example in the context of the group $H_3$ because of its practical relevance in the applied sciences. We then extend our analysis to the $H_2$ and $H_4$ cases in Section \ref{sec_other}, and conclude in Section \ref{concl} with a discussion of our formalism and the applications it opens up in quasicrystals, carbon chemistry and virology.}{}

\section{{Affine extensions of Coxeter groups via extensions of their Cartan matrices}}\label{sec_prelim}

In this section, we {recall} the representation of finite Coxeter groups in terms of root systems and show how  affine Coxeter groups can be derived  via an extension of their root systems and Cartan matrices. 

\subsection{Finite Coxeter groups and root systems} 

The elements of finite Coxeter groups can be visualised as reflections at planes through the origin in a Euclidean vector space $\mathcal{E}$. In particular, if $(\cdot , \cdot)$ denotes the inner product in $\mathcal{E}$, and $v$, $\alpha\in\mathcal{E}$, then 
\begin{equation}\label{reflect}
r_\alpha v = v - \frac{2(\alpha,  v)}{(\alpha, \alpha)}\alpha
\end{equation}
corresponds to  a Euclidean reflection at a plane perpendicular to $\alpha$.  
The structure of a finite Coxeter group can hence be encoded by a set of vectors $\Delta$, called the root system, that consists of all vectors normal to these reflection planes. Root systems fulfil the following properties: 

\begin{itemize}
\item If $\alpha\in \Delta$ and $\lambda \alpha \in \Delta$, then $\lambda = \pm 1$. 
\item $\Delta$ is invariant under all reflections $r_\alpha$ with $\alpha\in\Delta$. 
\end{itemize}

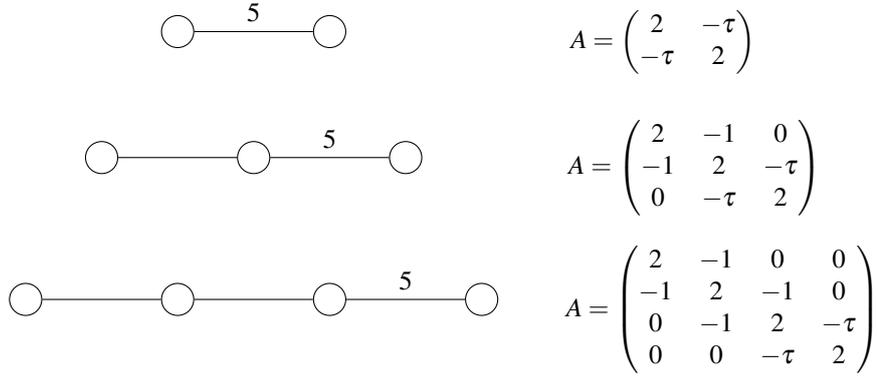
\begin{figure}
	\begin{center}
       \begin{tabular}{@{}c@{ }c@{ }}
		\begin{tikzpicture}[
		    knoten/.style={        circle,      inner sep=.15cm,        draw}  ]

		  \node at (1,1.5) (knoten1) [knoten,  color=white!0!black] {};
		  \node at (3,1.5) (knoten2) [knoten,  color=white!0!black] {};
		\node at (2,1.75)  (tau) {$5$};
		  \path  (knoten1) edge (knoten2);

		\end{tikzpicture}&

		$A = \begin{pmatrix} 2&-\tau\\ -\tau&2 \end{pmatrix}\label{CarH2}$

	\hspace{1cm}
	\\

		\begin{tikzpicture}[
		    knoten/.style={        circle,      inner sep=.15cm,        draw}  
		   ]

		  \node at (1,1.5) (knoten1) [knoten,  color=white!0!black] {};
		  \node at (3,1.5) (knoten2) [knoten,  color=white!0!black] {};
		  \node at (5,1.5) (knoten3) [knoten,  color=white!0!black] {};

		\node at (4,1.75)  (tau) {$5$};
	    \node at (4,2.5)   (tau1) {};

		  \path  (knoten1) edge (knoten2);
		  \path  (knoten2) edge (knoten3);
	\hspace{1cm}
\vspace{0.5cm}
		\end{tikzpicture} &

		\hspace{0.75cm}
		$	A = \begin{pmatrix} 2&-1&0 \\ -1&2&-\tau\\ 0&-\tau&2 \end{pmatrix}
			\label{CarH3}$

			\hspace{1cm}	 \\
 		\begin{tikzpicture}[
		    knoten/.style={        circle,      inner sep=.15cm,        draw}  
		   ]

		  \node at (1,1.5) (knoten1) [knoten,  color=white!0!black] {};
		  \node at (3,1.5) (knoten2) [knoten,  color=white!0!black] {};
		  \node at (5,1.5) (knoten3) [knoten,  color=white!0!black] {};
		  \node at (7,1.5) (knoten4) [knoten,  color=white!0!black] {};

		\node at (6,1.75)  (tau) {$5$};
	\node at (4,2.5)   (tau1) {};

		  \path  (knoten1) edge (knoten2);
		  \path  (knoten2) edge (knoten3);
		  \path  (knoten3) edge (knoten4);
\vspace{0.5cm}
		\end{tikzpicture}	

&

	\hspace{0.4cm}
	$	A = \begin{pmatrix} 2&-1&0&0 \\ -1&2&-1&0 \\ 0&-1&2&-\tau \\ 0&0&-\tau&2 \end{pmatrix}
		\label{CarH4}$

  \\
  \end{tabular}
  \caption[Hi]{Coxeter-Dynkin diagrams and Cartan matrices for, from top to bottom, $H_2$, $H_3$ and $H_4$. 
}
\label{figHi}
\end{center}
\end{figure}

For a crystallographic Coxeter group, a subset of $\Delta$, called simple roots, is sufficient to express every element of $\Delta$ via a $\mathbb{Z}$-linear combination with coefficients of the same sign. The set of vectors $\Delta$  is therefore completely characterised by the simple roots, which in turn completely characterise the Coxeter group. In the case of the non-crystallographic Coxeter groups $H_2$, $H_3$ and $H_4$, the same holds for the extended integer ring $\mathbb{Z}[\tau]=\lbrace a+\tau b| a,b \in \mathbb{Z}\rbrace$, where $\tau$ is   the golden ratio $\tau=\frac{1}{2}(1+\sqrt{5})$. Note that together with its  Galois conjugate $\tau'\equiv \sigma=\frac{1}{2}(1-\sqrt{5})$, $\tau$ satisfies the quadratic equation $x^2=x+1$.  The structure of the basis of simple roots is encoded by the Cartan matrix, which contains the geometrically invariant information of the root system as follows: 
\begin{equation}\label{CM}
	A_{ij}=2(\alpha_i, \alpha_j)/(\alpha_i, \alpha_i).
\end{equation}
We also indicate the relations of the root vectors in terms of Coxeter-Dynkin diagrams, where nodes correspond to simple roots and links labeled $m$ encode an angle of $\frac{\pi}{m}$ between the root vectors, with $m$ omitted if the angle is $\frac{\pi}{3}$ and no link shown if $\frac{\pi}{2}$. 
Coxeter-Dynkin diagrams and Cartan matrices for $H_2$, $H_3$ and $H_4$ are given in Fig. \ref{figHi}. 

\subsection{Affine extensions of finite Coxeter groups \label{aff_ext}}

For a Coxeter group with a crystallographic root system $\Delta$, an affine Coxeter group can be introduced as follows \cite{McCammond2010Coxeter}: For each $\alpha\in \Delta$ and $i\in\mathbb{Z}$, one defines \emph{affine hyperplanes} $H_{\alpha ,i}$ as solutions of the equations $(x,  \alpha) =i\,$.
The unique nontrivial isometry of $\mathcal{E}$ that fixes $H_{\alpha ,i}$ pointwise is called an affine reflection $r_{\alpha ,i}$. It corresponds to the reflections considered earlier if $i=0$. 

In the case of non-crystallographic Coxeter groups, this definition is not appropriate as $i\in \mathbb{Z}$ is not possible because the crystallographic restriction \cite{Senechal:1996}  implies that the planes cannot be stacked periodically; however, $i \in \mathbb{Z}[\tau]$ is too general because $\mathbb{Z}[\tau]$ is dense in $\mathbb{R}$.
But they could occur with spacings corresponding to  quasilattices.

Motivated by this relation with quasilattices, affinisations of the non-crystallographic Coxeter groups $H_2$, $H_3$ and $H_4$  have been studied in Ref.~\cite{Twarock:2002a}. In this work, the parallels with root systems in Lie algebras have been exploited and a formalism akin to the one used in Kac-Moody Theory has been employed to extend the root system so as to include roots that define affine reflections.

In particular, denoting by $\omega_j$ the basis of fundamental weights that is dual to the basis of simple roots $\alpha_i$, i.e. $\omega_j\alpha_i = \delta_{i,j}$, the Cartan matrix $A_{ij}$ corresponds to the coordinates of the root vectors in the basis of fundamental weights: 
\begin{equation}
\alpha_j = \sum_i A_{ji} \omega_i\,.
\end{equation}

{By extending the Cartan matrix by an additional row and column, we can} thus derive information on the root that defines the affine reflection. Following the Kac-Moody approach, extensions have been considered that abide by the following \emph{Kac-Moody-type extension rules}: 

\begin{enumerate}
\item  The diagonal entries in Eq.~(\ref{CM}) are normalised  as $A_{ii} = 2$.
\item  $A_{ij}\le 0$ and $A_{ij} = 0\Leftrightarrow A_{ji} = 0$ for $i\ne j$.
\item   Matrix entries are $\mathbb{Z}[\tau]$-valued.
\item  The affine extended matrix fulfils the \emph{determinant constraint} $\det A=0$. 
\end{enumerate}

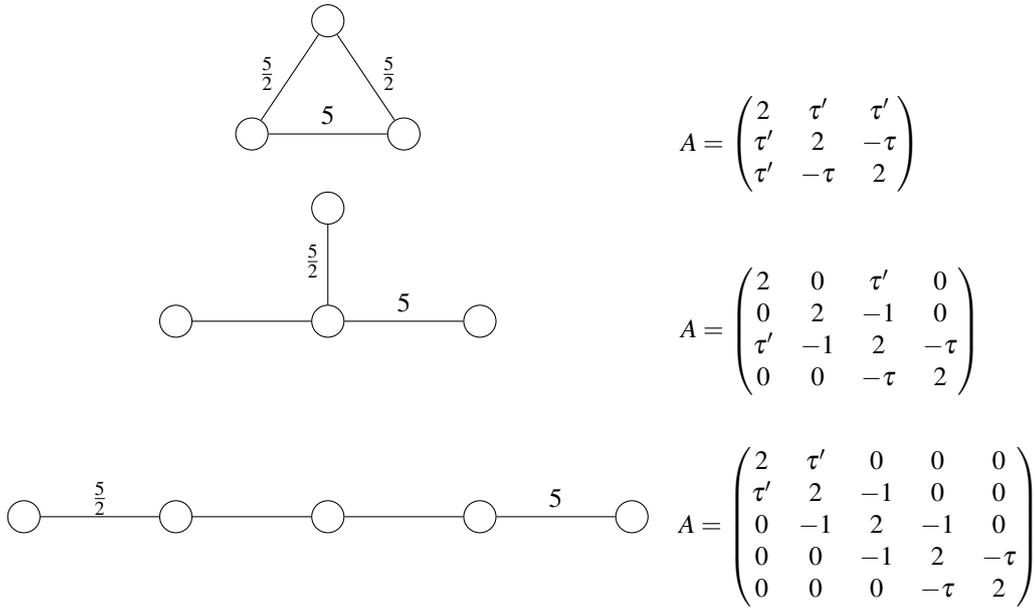
\begin{figure}
	\begin{center}
       \begin{tabular}{@{}c@{ }c@{ }}
		\begin{tikzpicture}[
		    knoten/.style={        circle,      inner sep=.15cm,        draw}  ]

			\node at (2,3) (knoten0) [knoten,  color=white!0!black] {};
			  \node at (1,1.5) (knoten1) [knoten,  color=white!0!black] {};
			  \node at (3,1.5) (knoten2) [knoten,  color=white!0!black] {};
			  \node at (1.2,2.3)  (sigma) {$\frac{5}{2}$};
			  \node at (2.8,2.3)  (sigma) {$\frac{5}{2}$};
			  \node at (2,1.75)  (tau) {$5$};
			  \path  (knoten1) edge (knoten2);
			  \path  (knoten1) edge (knoten0);
			  \path  (knoten0) edge (knoten2);

		\end{tikzpicture}&
		
		\hspace{-0.55cm}

		$A = \begin{pmatrix} 2&\tau'&\tau' \\ \tau'&2&-\tau\\ \tau'&-\tau&2 \end{pmatrix}
		\label{CarH2aff}$

	\hspace{1cm}
	\\
		
		\begin{tikzpicture}[
		    knoten/.style={        circle,      inner sep=.15cm,        draw}  
		   ]

		  \node at (3,3) (knoten0) [knoten,  color=white!0!black] {};  
		  \node at (1,1.5) (knoten1) [knoten,  color=white!0!black] {};
		  \node at (3,1.5) (knoten2) [knoten,  color=white!0!black] {};
		  \node at (5,1.5) (knoten3) [knoten,  color=white!0!black] {};
		\node at (4,1.75)  (tau) {$5$};
		\node at (2.8,2.3)  (sigma) {$\frac{5}{2}$};
		  \path  (knoten1) edge (knoten2);
		  \path  (knoten2) edge (knoten3);
		  \path  (knoten0) edge (knoten2);
	\hspace{1cm}
\vspace{0.5cm}
		\end{tikzpicture} &
		
		\hspace{0.25cm}
		$A = \begin{pmatrix} 2&0&\tau'&0 \\ 0&2&-1&0 \\ \tau'&-1&2&-\tau\\ 0&0&-\tau&2 \end{pmatrix}
		\label{CarH3aff}$
			\hspace{1cm}	 \\
 		\begin{tikzpicture}[
		    knoten/.style={        circle,      inner sep=.15cm,        draw}  
		   ]

		 \node at (-1,1.5) (knoten0) [knoten,  color=white!0!black] {};  
		  \node at (1,1.5) (knoten1) [knoten,  color=white!0!black] {};
		  \node at (3,1.5) (knoten2) [knoten,  color=white!0!black] {};
		  \node at (5,1.5) (knoten3) [knoten,  color=white!0!black] {};
		  \node at (7,1.5) (knoten4) [knoten,  color=white!0!black] {};

		\node at (0,1.75)  (tau) {$\frac{5}{2}$};
		\node at (6,1.75)  (tau) {$5$};

		\node at (4,3)  (tau2) {};

		  \path  (knoten0) edge (knoten1);
		  \path  (knoten1) edge (knoten2);
		  \path  (knoten2) edge (knoten3);
		  \path  (knoten3) edge (knoten4);
\vspace{0.5cm}
		\end{tikzpicture}	

&

	$	A = \begin{pmatrix} 2&\tau'&0&0&0 \\ \tau'&2&-1&0&0 \\ 0&-1&2&-1&0 \\ 0&0&-1&2&-\tau \\ 0&0&0&-\tau&2 \end{pmatrix}\label{CarH4aff}$

  \\
  \end{tabular}
  \caption[Hiaff]{Coxeter-Dynkin diagrams and Cartan matrices for  $H_2^{aff}$, $H_3^{aff}$ and $H_4^{aff}$, the unique symmetric affine extensions in Ref.~\cite{Twarock:2002a}. 
Note that  angles of $2\pi/5$ lead to labels $\frac{5}{2}$ (or, $\tau'$ in the notation of Ref.~\cite{Twarock:2002a}).}
\label{figHiaff}
\end{center}
\end{figure}

It has been shown in Ref.~\cite{Twarock:2002a} that there is a unique affine extension for $H_2$, $H_3$ and $H_4$ if the extended matrix is required to be symmetric.  In each case, the affine root can be expressed as $\alpha_0=-\alpha_H$, where $\alpha_H$ corresponds to the highest root in the root system; the corresponding Coxeter-Dynkin diagrams and Cartan matrices are given in  Fig.~\ref{figHiaff}.

On the group level, these extensions correspond to the addition of the following affine reflection to the group: 
\begin{equation}\label{reflectaff}
r^{aff}_\alpha v = \alpha + v - \frac{2(\alpha,v)}{(\alpha,\alpha)}\alpha,
\end{equation} 
where $\alpha\equiv \alpha_H$.

Together with the reflection $r_\alpha\equiv r_{\alpha_H}$ from $H_i$, it generates a translation $T$ as follows: 
\begin{equation}	
r_\alpha^{aff}r_\alpha v = r_\alpha^{aff} \left( v - \frac{2(\alpha,  v)}{(\alpha, \alpha)}\alpha\right)=
\alpha+v=: T v.
\label{Transl}
\end{equation}

In the context of Ref.~\cite{Twarock:2002a}, the affine reflection thus gives rise to one specific translation. Note that from a geometric point of view, Eq. (\ref{reflectaff}) and Eq. (\ref{Transl}) are independent of whether $\alpha$ is a root vector, because any translation of length $2l$ can be generated by two parallel reflection planes separated by a distance $l$.  Thus,  substituting $\alpha$ by $\lambda \alpha$ in Eq. (\ref{reflectaff}) leads to a translation of length $\lambda T$ along the direction of $\alpha$.  This relation of parallel reflection planes with translations will be crucial later.

\section{{New affine extensions of non-crystallographic Coxeter groups}}\label{sec_new}

{We  investigate here asymmetric}  extensions of non-crystallographic Coxeter groups. These have not been considered before, while -- as we will demonstrate in this paper --  they  lead to interesting  applications, most notably in virology. 
As in Ref.~\cite{Twarock:2002a}, we consider Kac-Moody-type extensions of $H_2$, $H_3$ and $H_4$, but allow for asymmetric Cartan matrices with entries in  $\mathbb{Z}[\tau]$. Due to the interpretation of the Cartan matrix in terms of scalar products between root vectors, there are constraints on such extensions that we formulate in a  Lemma.


\textbf{Lemma}: In the case of a Coxeter group with a simply-laced Coxeter-Dynkin diagram, a Cartan matrix extended according to the Kac-Moody-type extension rules fulfils
\begin{equation}
\frac{A_{i0}}{A_{0i}}=\frac{A_{k0}}{A_{0k}}\,\,\, \forall i, k \text{ with } A_{0i}, A_{0k}\ne 0.\label{lemma1SL}
\end{equation}

\emph{Proof:}
Since the entries of the Cartan matrix of a root system are given in terms of the lengths of the roots $l^2_j:=\left(\alpha_j, \alpha_j\right)=\left|\alpha_j\right|^2$ and the angles  $\theta_{ij}$ between roots $\alpha_i$ and $\alpha_j$ by  (no summation implied)
\begin{equation}
A_{ij} =2\frac{\left(\alpha_i, \alpha_j\right)}{\left(\alpha_i, \alpha_i\right)}=2\frac{\left|\alpha_j\right|}{\left|\alpha_i\right|}\cos  \theta_{ij},\label{Cartanmatrix}
\end{equation}
it follows  that
\begin{equation}
l^2_j=\frac{A_{ij}}{A_{ji}}l^2_i\label{length}
\end{equation}
 and 	
\begin{equation}
\cos ^2 \theta_{ij}=\frac{1}{4}A_{ij}A_{ji}.\label{angle}
\end{equation}

When $i=k$ for some non-zero entry $A_{0k}$ in  Eq.~(\ref{length}), the length of the new root $\alpha_0$ is  given by $l^2_0=\frac{A_{k0}}{A_{0k}}l^2_k$. The possible non-zero inner products of the remaining roots $\alpha_i$ with the new root are therefore constrained by
$\frac{A_{i0}}{A_{0i}}=\frac{A_{k0}}{A_{0k}}\frac{l^2_k}{l^2_i}$. 
In the simply-laced case $l_i=l_k$ and the claim follows.  

This consistency condition on the additional root severely restricts the  non-zero entries in the Cartan matrices, particularly in the integer-valued setting of simple Lie algebras, because one only needs to  consider one pair and several multipliers  e.g. $A_{i0}=\beta_i A_{k0}, A_{0i}=\beta_i A_{0k}$.  In particular, this implies the following 

\textbf{Corollary}:  If there are any entries $A_{0k},A_{k0}$  with $A_{0k}=A_{k0}$ in the extended Cartan matrix, then it must be totally symmetric. 

\emph{Proof:}
$A_{0k}=A_{k0}$ implies $\frac{A_{k0}}{A_{0k}}=1$  in  Eq.~(\ref{lemma1SL}), hence
$A_{i0}=A_{0i}$ for any $i$. 

The Corollary says that if one of the simple roots in a simply-laced system (before extension) has the same length as the new root, then so do all others. {For $H_2$, $H_3$ and $H_4$ this case is known, c.f. Ref.~\cite{Twarock:2002a} in the non-crystallographic and Ref.~\cite{FuchsSchweigert:1997} in the crystallographic case. } In contrast, we consider here the case of asymmetric affine extensions.  

Note that in  Lie Theory, which covers the (crystallographic) Weyl groups, Eq.~(\ref{angle}) severely restricts the possible (integer) combinations of entries that can occur in the Cartan matrix, as $0\le \cos ^2\theta \le 1$ leads to the constraint
\begin{equation}
A_{ij}A_{ji}=4\cos ^2 \theta_{ij}\le 4.\label{angleconstraint}
\end{equation}
 However, this condition is not nearly as restrictive in the non-crystallographic setting, as $\mathbb{Z}[\tau]$ is dense in $\mathbb{R}$.

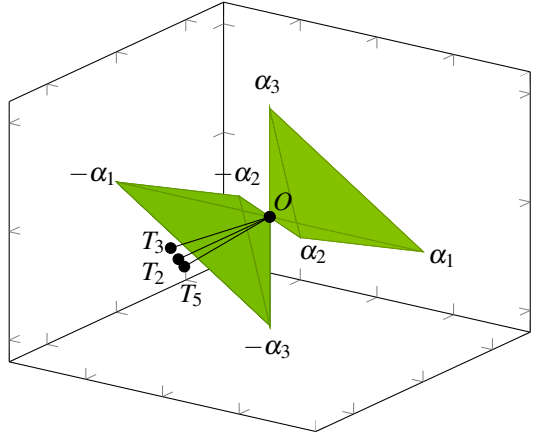
\begin{figure}
	\begin{center}


\begin{tikzpicture}
		\begin{axis}[
			xticklabels={,,}, yticklabels={,,}, zticklabels={,,},
			view={-235}{25},colormap/greenyellow]
 
\addplot3[surf,mesh/rows=3] coordinates { 
(0,0,0) (0,1,0) (-0.5*0.618,-0.5,-0.5*1.618)
(0,0,0) (0,1,0) (0,0,1)
(0,0,0) (0,0,1) (-0.5*0.618,-0.5,-0.5*1.618) 
(0,0,1) (-0.5*0.618,-0.5,-0.5*1.618) (0,1,0)
(0,1,0) (0,0,1) (0,0,0) 
}; 

\addplot3[surf,mesh/rows=3] coordinates { 
(0,0,0) (0,-1,0) (0.5*0.618,0.5,0.5*1.618)
(0,0,0) (0,-1,0) (0,0,-1)
(0,0,0) (0,0,-1) (0.5*0.618,0.5,0.5*1.618) 
(0,0,-1) (0.5*0.618,0.5,0.5*1.618) (0,-1,0)
(0,-1,0) (0,0,-1) (0,0,0) 
};

\addplot3 [mark=*] coordinates {(0,0,0) (1/3.8,0,0) };
\addplot3 [mark=*] coordinates {(0,0,0) (1.618/1.73/3.8,0,-0.618/1.73/3.8)  };
\addplot3 [mark=*] coordinates {(0,0,0) (1.618/1.9/3.8, -1/1.9/3.8, 0) };

\end{axis} 
\node at (3.4,4.6)  (olabel)  {$\alpha_3$};
\node at (3.4,1.1)  (olabel)  {$-\alpha_3$};	
\node at (3.6,3.05)  (olabel)  {$O$};
\node at (5.7,2.3)  (olabel)  {$\alpha_1$};	
\node at (1.1, 3.4)  (olabel)  {$-\alpha_1$};	
\node at (4.0,2.35)  (olabel)  {$\alpha_2$};
\node at (3.0,3.4)  (olabel)  {$-\alpha_2$};
\node at (2.4,1.8)  (olabel)  {$T_5$};
\node at (1.9,2.5)  (olabel)  {$T_3$};
\node at (1.9,2.1)  (olabel)  {$T_2$};
\end{tikzpicture}

\caption[dummy1]{Geometry of the opening cone.}
\label{fig_cone}
\end{center}
\end{figure}

Motivated by applications in virology \cite{Keef:2009}, we  determine here extended Cartan matrices that encode affine reflections perpendicular to 2-, 3- or 5-fold axes of icosahedral symmetry. We will use the following simple roots for $H_3$:
\begin{equation}
\alpha_1=(0,1,0),\,\,\,\alpha_2=-\frac{1}{2}(-\sigma,1,\tau),\,\,\,\alpha_3=(0,0,1).\label{simpleroots}
\end{equation}
The unique 2-fold, 3-fold and 5-fold symmetry axes lying in the opening cone of $-\alpha_1$, $-\alpha_2$ and $-\alpha_3$, i.e. having only negative-valued (or zero) scalar products with the basis of simple roots, are then given by
\begin{equation}
T_2=(1,0,0),\,\,\,T_3=(\tau,0,  \sigma),\,\,\,T_5=(\tau, -1,0).\label{simplerootscone}
\end{equation}
This is important in the following as every other choice of axes would violate the negativity requirements of the Cartan matrices, although they would still satisfy the consistency Lemma.
This geometry is illustrated in Fig. \ref{fig_cone}.

\subsection{Affine extensions of $H_3$ by reflection planes perpendicular to two-fold axes} \label{sec_2fold}

Due to the geometry of the root system of $H_3$,
an affine root $\alpha_0$  parallel to $T_2$ is  orthogonal to $\alpha_1$ and $\alpha_3$, and hence the only non-vanishing inner product is with $\alpha_2$.
The length of  $\alpha_0$ is then  given by the degree of asymmetry of the Cartan matrix. In particular, an $\alpha_0$ of any  length not equal to that of the simple roots \emph{requires} the  extended Cartan matrix to be   asymmetric. 

Thus, for affine reflection planes perpendicular to $T_2$ in  Eq.~(\ref{simplerootscone}), we get a family of Cartan matrices of the form
\begin{equation}
A = \begin{pmatrix} 2&0&x &0 \\ 0&2&-1&0 \\ y&-1&2&-\tau\\ 0&0&-\tau&2 \end{pmatrix},\label{CarH3affASfam2}
\end{equation}
where we need to stipulate  $xy=\sigma^2=2-\tau$ in order to enforce that the determinant vanishes, i.e. for the extension to be affine. We note that we  recover the results of  Ref.~\cite{Twarock:2002a}, i.e. for $H_3^{aff}$, if $x=y=\sigma$. 

We now solve the equation 
\begin{align}
xy=2-\tau=\sigma^2,  \,\,\, x,y \in \mathbb{Z}[\tau]\label{xy}
\end{align}
by writing $x=(a+\tau b)$ and $y=(c+\tau d)$ with $a,b,c,d \in \mathbb{Z}$. Using $xy=A_{ij}A_{ji}$ and  Eq.~(\ref{angle}) we infer the angles of the affine root with the simple roots,  confirming that  matrices as in Eq.~(\ref{CarH3affASfam2}) all correspond to affine roots along $T_2$.  {Note that the constraint from Eq.~(\ref{angleconstraint}) is weaker than the determinant constraint and hence need not be considered.} 

 The affine extension $H_3^{aff}$ found in Ref.~\cite{Twarock:2002a} corresponds to the simplest case 
$(a,b;c,d)=(1, -1; 1, -1)$, i.e. to $x=y=\sigma$. As the  (non-trivial) units in  $\mathbb{Z}[\tau]$ are  $\tau^k,\, k\in  \mathbb{Z}$,  we can generate a series of solutions to the determinant constraint by scaling $x\rightarrow \tau^{-k} x$ and $y\rightarrow \tau^k y$, as this leaves the product $xy$ invariant.  Since these are the only units in  $\mathbb{Z}[\tau]$, these rescaled versions are in fact the only additional solutions to the determinant constraint. In terms of $(a,b;c,d)$, this amounts  to the substitutions
\begin{equation*}\label{Fibscal1}
 (a,b;c,d)\rightarrow (b, a+b; d-c, c) \text{ for multiplication by }(\tau, \tau^{-1}) \text{, and to \\ }
\end{equation*}
\begin{equation}\label{Fibscal2}
(a,b;c,d)\rightarrow (b-a, a; d, c+d) \text{ for multiplication by }(\tau^{-1}, \tau). 
\end{equation}
The quadruplet $(-2, 1;-1, 0)$ corresponding to the solution $(x,y)=(-\tau^{-2}, -1)$ is the solution with the smallest value of  $\sum |a_i|+|b_i|+|c_i|+|d_i|$ and is in this sense the start of the asymmetric series of solutions generated by the rescalings in  Eq.~(\ref{Fibscal2}). Note that the transformations in Eq.~(\ref{Fibscal2})  act on each pair $(a,b)$ and $(c,d)$ as
\begin{equation}
\begin{pmatrix} a' \\b'  \end{pmatrix}=\begin{pmatrix} 0&1\\ 1&1 \end{pmatrix}\begin{pmatrix} a \\b  \end{pmatrix},\label{Fib}
\end{equation}
which is the substitution matrix generating the Fibonacci series. 
We therefore call this family of solutions the \emph{Fibonacci family}. 

The length of the affine root is determined according to  Eq.~(\ref{length}) by $\sqrt{{x}/{y}}$, where we have used the fact that all roots of the unextended root system are considered to have length $1$. In particular, for the symmetric case $x=y$, the affine root has the same length as the roots in $\Delta$, as expected from  Ref.~\cite{Twarock:2002a}. 

Since $x$ and $y$ commute in  Eq.~(\ref{xy}), considering  $(c,d;a,b)$ instead of $(a,b;c,d)$ trivially generates another solution to the determinant constraint. However, 
the root length is determined by the quotient $\sqrt{{x}/{y}}$. Therefore, swapping $x$ and $y$ generates a solution of different length.  Thus, for generic determinant constraints, one can generate two Fibonacci families of solutions by $\tau$-multiplication and swapping factors.  Here, the determinant constraint has a symmetric solution given by the  quadruplet $(a,b;c,d)=(1, -1; 1, -1)$, so that swapping does not in fact yield anything new. Thus, in this case one only has  one family indexed by powers of $\tau$. Later examples will have two independent solutions, however. 

Correspondingly, for the Fibonacci family of solutions obtained via the scalings $x\rightarrow \tau^{k} x$ and $y\rightarrow \tau^{-k} y$, the affine roots have lengths $\tau^{k}$. Via the argument presented in Section \ref{aff_ext},  this affine root defines a new reflection plane at the group level that does not go through the origin, but is parallel to one of the reflection planes in $H_3$. 
As a consequence of  Eq.~(\ref{Transl}), one obtains a translation  with a translation length determined by the length of the affine root.
 The affine roots determined here therefore generate translations of lengths $\tau^{k}$, which contain the five translation lengths listed in  Ref.~\cite{Keef:2009}  $\{\tau'^2, -\tau', 1, \tau, \tau^2\}=\{\tau^{-2}, \tau^{-1}, 1, \tau, \tau^2\}$. 
 
 We hence obtain a countably infinite set ($k\in \mathbb{Z}$) of  affine extensions of $H_3$ with affine reflection planes at  distances  $\tau^{k}/2$ from the origin. For any given $k$, there is  an infinite stack of parallel planes with separation $\tau^{k}/2$.

\subsection{Affine extensions of $H_3$ by reflection planes perpendicular to three-fold axes} 

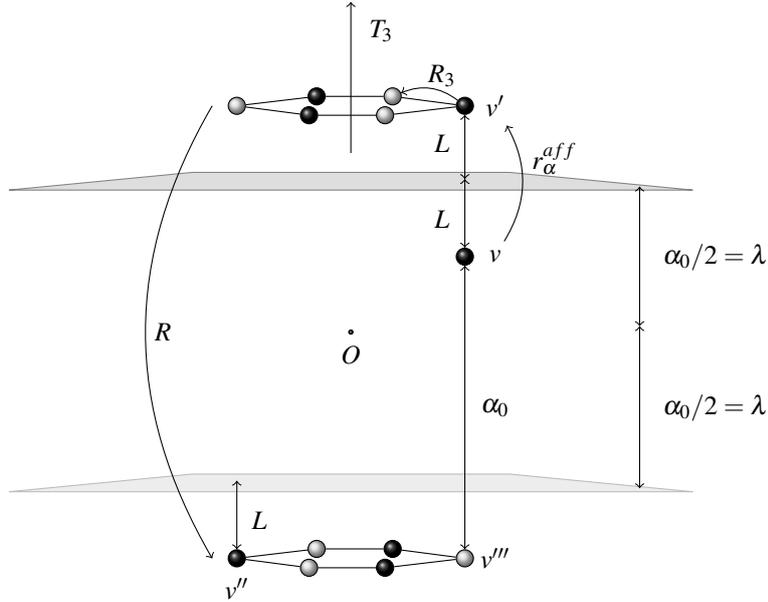
\begin{figure}
\begin{center}
\tikzstyle{background grid}=[draw, black!50,step=.5cm]

		\begin{tikzpicture}[
		    knoten/.style={        circle,      inner sep=.08cm,        draw}  , 
		   dot/.style={        circle,      inner sep=.02cm,        draw}  , 
		my node/.style={trapezium, fill=#1!20, draw=#1!75, text=black} 
		   ]
		
		\node [my node=gray!70!black, minimum width=9cm, trapezium stretches] 
		at (0,2) {}; 
		\node [my node=gray!70!white, minimum width=9cm, trapezium stretches] 
		at (0,-2) {};

		  \node at (-1.5, 3)   (hex1)  [knoten,  ball color=gray!30!white] {};
		  \node at (1.5, 3) (hex2) [knoten, ball color=black] {};  
		  \node at (-0.5+0.05, 3.125) (hex3) [knoten, ball color=black] {};
		  \node at (0.5+0.05, 3.125) (hex4) [knoten,  ball color=gray!30!white] {};
		  \node at (-0.5-0.05, 2.875) (hex5) [knoten, ball color=black] {};
		  \node at (0.5-0.05, 2.875) (hex6) [knoten, ball color=gray!30!white] {};
		  \path  (hex1) edge (hex3);
		  \path  (hex1) edge (hex5);
		  \path  (hex2) edge (hex4);
		  \path  (hex2) edge (hex6);
		  \path  (hex3) edge (hex4);
		  \path  (hex5) edge (hex6);
		  \node at (1.5, 1) (vnode) [knoten, ball color=black] {};  		
	      \node at (1.9,1)  (v) {$v$};
		  \node at (1.9,3)  (vp) {$v'$};
		  \node at (1.5,1.95)  (raff) {};
		  \path [<->] (raff.mid) edge (vnode);
		  \path [<->] (raff.mid) edge (hex2);
		  \node at (1.2,1.5)  (vL) {$L$};
		  \node at (1.2,2.5)  (vpL) {$L$};
	      \node at (0,0)  (o) [dot, ball color=black!100!white] {};
	      \node at (0,-0.3)  (olabel)  {$O$};	
			
		  \node at (0,2.25)  (T3beg) {};		
		  \node at (0,4.5)  (T3end) {};	
		  \node at (0.4,4)  (T3label) {$T_3$};
		  \path [->] (T3beg) edge (T3end);		
	
		  \node at (-1.5, -3)   (ahex1)[knoten, ball color=black] {}; 
		  \node at (1.5, -3) (ahex2)[knoten,  ball color=gray!30!white] {};
		  \node at (-0.5+0.05, -2.875) (ahex3) [knoten,  ball color=gray!30!white] {};
		  \node at (0.5+0.05, -2.875) (ahex4) [knoten, ball color=black] {}; 
		  \node at (-0.5-0.05, -3.125) (ahex5)[knoten,  ball color=gray!30!white] {};
		  \node at (0.5-0.05, -3.125) (ahex6) [knoten, ball color=black] {}; 
		  \path  (ahex1) edge (ahex3);
		  \path  (ahex1) edge (ahex5);
		  \path  (ahex2) edge (ahex4);
		  \path  (ahex2) edge (ahex6);
		  \path  (ahex3) edge (ahex4);
		  \path  (ahex5) edge (ahex6);
	      \node at (-1.5,-3.4)  (vpp) {$v''$};
	      \node at (1.9,-3)  (vppp) {$v'''$};
		  \path [<->] (vnode) edge (ahex2);
		  \node at (1.9,-1)  (va0) {$\alpha_0$};
		  \node at (-1.5,-2.05)  (raff2) {};		
		  \node at (-1.2,-2.5)  (vppL) {$L$};
		  \node at (-1.7, 3)   (hex1b) {};		
	      \node at (-1.7, -3)   (ahex1b) {}; 		
		  \path [<->] (raff2.mid) edge (ahex1);
		  \path[->] (hex1b.west) edge [bend right] node [right] {$R$} (ahex1b.west) ;
		  \path[->] (hex2) edge [bend right] (hex4) ;
	      \node at (1.2, 3.4)   (R3) {$R_3$}; 	
		  \node at (3.8,-2.15)  (raff2b) {};
		  \node at (3.8,1.85)  (raffb) {};
		  \node at (4.8,-1)  (raff2c) {$\alpha_0/2=\lambda$};
		  \node at (4.8,1)  (raffc) {$\alpha_0/2=\lambda$};		
		  \node at (3.8,0)  (ob) {};	
	      \path [<->] (ob.mid) edge (raffb.mid);			
	      \path [<->] (ob.mid) edge (raff2b.mid);
		  \path[->] (v) edge [bend right] node [above right] {$r_\alpha^{aff}$} (vp) ;		

	\hspace{1cm}
\vspace{0.5cm}
		\end{tikzpicture}

\end{center}

\caption[dummy1]{Translations and twist translations can be generated by affine reflection planes perpendicular to 3- and 5-fold rotational symmetry axes of icosahedral symmetry. Assume that there is  an affine reflection plane at a distance $\lambda=\frac{\alpha_0}{2}$ from the origin (but no corresponding parallel plane through the origin). Any vector $v$ at a distance $L$ below the affine reflection plane can be reflected to a vector $v'$ at a position $L$ above it. The rotational part of the group ($R_3$ and $R_5$) then generates a triangle around the 3-fold axis (black points), or a pentagon around a 5-fold axis, and the full group $H_3$ generates a hexagon (additional white points), or a decagon, respectively. By using a rotation $R$ in the group, these can  be rotated into a parallel plane at a distance $2\lambda+L$ from the affine reflection plane and  at a distance $\lambda+L$ from the origin (note that $v''$ is part of this plane). As the generators $R_3$ and $R_5$ allow to rotate any point in the  hexagon and  decagon into any other, one of these rotations can be used to align the reflected and rotated $v$ (the point $v''$) with the initial $v$ (yielding a point $v'''$) such that  $v$ and $v'''$ differ by  a translation of length $2\lambda=\alpha_0$ along $T_3$ (resp. $T_5$). The other 5 (9 respectively) rotations also generate twist translations. }
\label{figtwist}
\end{figure}

We next consider affine roots parallel to $T_3$ in  Eq.~(\ref{simplerootscone}).  These are by construction 
 orthogonal to the simple roots $\alpha_1$ and $\alpha_2$, so that only the  inner product with $\alpha_3$ gives non-zero entries in the Cartan matrix:
\begin{equation}
A = \begin{pmatrix} 2&0&0 &x \\ 0&2&-1&0 \\ 0&-1&2&-\tau\\ y&0&-\tau&2 \end{pmatrix}.\label{CarH3affASfam3}
\end{equation}
This time the determinant constraint
$xy=\frac{4}{3}\sigma^2=\frac{4}{3}(2-\tau)$ is no longer $\mathbb{Z}[\tau]$-valued, and hence solutions  do not exist in $\mathbb{Z}[\tau]$. However, we can still entertain the idea of affine reflection planes perpendicular to 3-fold axes, which generate translations and twist translations along $T_3$ as we demonstrate in Fig. \ref{figtwist}. 

In particular, for a vector $v$ as in the figure, reflection $r_\alpha ^{aff}$ at the affine plane plus subsequent rotation $R$ (to $v''$ in the figure) yields a twist translation 
\begin{equation}
T^{twist}v =Rr_\alpha^{aff} =-{\alpha_0}+R_3^kv,\label{twist}
\end{equation}
where $\alpha_0$ is the affine root and $R_3^k$ denotes a rotation by $\frac{\pi k}{3}, \,\, 0\le k\le5, k\in \mathbb{Z}$ around $T_3$. Moreover, using $R_3^l$ with $k+l=6$, we obtain a translation
\begin{equation}
 Tv=R_3^lRr_\alpha^{aff}v=v-{\alpha_0}.\label{twtrans}
\end{equation}

In order to find  affine extensions of the type in Eq. (\ref{CarH3affASfam3}), we introduce $\gamma, \delta \in\mathbb{Q}$ as  normal fractions and write the entries of the Cartan matrix as a  $\mathbb{Z}[\tau]$-integer multiplied by $\gamma, \delta$ as $x=\gamma(a+\tau b)$ and $y=\delta(c+\tau d)$. The   solution is therefore now given by a Fibonacci quadruplet $(a,b;c,d)$ plus multipliers $(\gamma, \delta)$, and the latter need to be such that they account for the fraction in the determinant constraint, e.g. $\gamma\delta=\frac{4}{3}$ in  Eq.~(\ref{CarH3affASfam3}). Thus the  root lengths are  $\sqrt{{x}/{y}}=\sqrt{{\gamma}/{\delta}}\tau^k=\gamma\sqrt{1/{\gamma\delta}}\tau^k=\sqrt{3}/2\gamma\tau^k$. 
For instance $x=\frac{3}{4}(1-\tau)$ and $y=(1-\tau)$ would be a solution of length $\frac{1}{2}\sqrt{3}$. 

As $T_3$ is  of length $\sqrt{3}$, this example corresponds to $\alpha_0=\frac{1}{2}T_3$, which induces a whole family of roots   $\alpha_0^k=\frac{1}{2}\tau^kT_3$. However, $x=(1-\tau)$ and $y=\frac{3}{4}(1-\tau)$ would also be a  solution, with the length $\frac{2}{\sqrt{3}}=\frac{2}{3}\sqrt{3}$, i.e.  $\alpha_0=\frac{2}{3}T_3$, with a family  $\alpha_0^k=\frac{2}{3}\tau^kT_3$. For  $x=\frac{1}{4}(1-\tau)$ and $y=3(1-\tau)$ we get length $\frac{1}{6}\sqrt{3}$, i.e. of roots $\alpha_0^k=\frac{1}{6}\tau^kT_3$. On the group level, there is a series of solutions for each choice of $\gamma$. In a vector space representation, Refs ~\cite{Wardman2011Thesis, Keef:2009} have found sets of distinguished translations in the context of extending  an icosidodecahedron (the root system of $H_3$) by translations or twist-translations along a 3-fold axis. 
For this special setting, the three series above for the choices $\gamma=\frac{3}{4}, 1, \frac{1}{4}$   and by allowing for linear combinations among the resulting translations, give all the results in Refs ~\cite{Wardman2011Thesis, Keef:2009}.
In particular, in Ref. ~\cite{Keef:2009} only instances of the first family were found, corresponding to lengths $\{\frac{1}{2}, \frac{1}{2}\tau, \frac{1}{2}\tau^2\}$, and $\{ 1, \tau^2\}$ respectively. Note that the latter are also implicitly contained in the former, corresponding to even multiples of these translations. 

In the previous section,  in the context of extensions along 2-fold axes, we were only searching for integer solutions.
  However,  the considerations in this section motivate allowing for $\mathbb{Q}[\tau]$-valued entries in the Cartan matrix such that $\gamma\delta=1$, where we  define $\mathbb{Q}[\tau]=\lbrace a+\tau b| a,b \in \mathbb{Q}\rbrace$. It is  interesting to analyse the translation lengths for the 2-fold axis in this more general setting, giving lengths $\sqrt{\frac{x}{y}}=\sqrt{\frac{\gamma}{\delta}}\tau^k=\gamma \tau^k$. $\gamma=1$ recovers the 5 solutions in $\mathbb{Z}[\tau]$  corresponding to the translations  $\{\tau^{-2}, \tau^{-1}, 1, \tau, \tau^2\}$ from Ref.~\cite{Keef:2009} listed earlier in Section \ref{sec_2fold}. $\gamma=\frac{1}{2}$ and $\gamma=\frac{3}{2}$ yield the 7 translations of length $\frac{1}{2}\{\tau^{-3},\tau^{-2}, \tau^{-1}, 1, \tau, \tau^2, \tau^3\}$ and the three translations  of lengths $\frac{3}{2}\{\tau^{-1}, 1, \tau\}$ that were found in Ref.~\cite{Wardman2011Thesis} for extensions of an icosidodecahedron along a 2-fold axis. In fact, by allowing linear combinations amongst the translations in those three families, we can accommodate all 26 translations listed in Ref.~\cite{Wardman2011Thesis} in a Coxeter group framework. Thus, in order to explain the results in the more general setting of Ref.~\cite{Wardman2011Thesis}, one in fact has to allow for  solutions in $\mathbb{Q}[\tau]$ even for the 2-fold axes.

\subsection{Affine extensions of $H_3$ by reflection planes perpendicular to five-fold axes} 

Considering now  affine roots parallel to $T_5$, we get a similar family of matrices of the form
\begin{equation}
A = \begin{pmatrix} 2&x&0 &0 \\ y&2&-1&0 \\ 0&-1&2&-\tau\\ 0&0&-\tau&2 \end{pmatrix}, \label{CarH3affASfam1}
\end{equation}
as again the 5-fold axis is perpendicular to two of the simple roots. This time, 
 the determinant constraint is $xy=\frac{4}{5}(3-\tau)$. We have dealt with the problem of the quotient in the previous section. However, here another new phenomenon occurs in that $(3-\tau)$ no longer has a symmetric solution in the integers. It has the obvious solutions $(-3, 1; -1, 0)$ and  $(-1, 0;-3, 1 )$, which are now independent, and hence generate  two different series (in terms of Fibonacci and Lucas numbers) by $\tau$-multiplication.
Since the determinant constraint implies a unique non-trivial angle for the new root with the simple roots (see  Eq.~(\ref{angle})), all members of the above family are indeed along $T_5$.

In the context of concrete applications in virology and carbon chemistry,  Refs \cite{Wardman2011Thesis, Keef:2009} have found distinguished translation lengths for extending icosahedrally symmetric polytopes along a 5-fold axis. We now show how all their results can be easily accommodated here. 
The  lengths for  an affine root corresponding to  $(-3, 1; -1, 0)$  are given by $\sqrt{\frac{5}{4}(3-\tau)}\gamma\tau^k$.  Via the  argument in the preceding section,  the affine root generates a translation or twist translation at the group level, c.f. the caption of Fig. \ref{figtwist}. Since the length of $T_5$ is $\sqrt{\tau+2}$, and $(3-\tau)\tau^2=(2+\tau)$, the resulting lengths are $\sqrt{\frac{5}{4}(2+\tau)}\gamma\tau^{k-1}$, or $\alpha_0^k=\sqrt{\frac{5}{4}}\gamma\tau^{k-1}T_5$. 

Because of the identity  $(3-\tau)(2+\tau)=5$, the corresponding result $\sqrt{\frac{5}{4(3-\tau)}}\gamma\tau^k$ for the second series generated by $( -1, 0; -3, 1)$ simplifies to $\alpha_0^k={\frac{1}{2}}\gamma\tau^{k}T_5$.
Thus, for instance, the choice $\gamma=1$ in the second family generates lengths $\{\frac{1}{2} \tau^{-1}, \frac{1}{2}, \frac{1}{2}\tau,\frac{1}{2} \tau^2\}$, which induces four of the results listed in  Ref.~\cite{Wardman2011Thesis}.  Ref.~\cite{Keef:2009} contains two further cases  with translations of length $\{1, \tau\}$, which correspond to  even multiples of the $\gamma=1$ translations, or are contained in the $\gamma=2$ series.

Note that a translation of the same length of $\frac{1}{2}$ could be generated in the first family by setting $\gamma=1/\sqrt{5}$. Making the same choice $\gamma=1/\sqrt{5}$ for the second family yields length ${\frac{1}{\sqrt{20}}}\tau^{k}T_5$, which due to the identity ${\frac{1}{\sqrt{20}}}\tau={\frac{1}{{10}}}(2+\tau)$ are equivalent to 
translation lengths  of the form $\frac{1}{10}(\tau+2)\tau^{k-1}$. Together with  $\frac{1}{5}(\tau+2)\tau^{k-1}$ and $\frac{2}{5}(\tau+2)\tau^{k-1}$, it  corresponds to most of the  results in Ref.~\cite{Wardman2011Thesis}:   $\frac{1}{10}(2+\tau)\{\tau^{-3},\tau^{-2}, \tau, \tau^{2}\}$,  $\frac{1}{5}(2+\tau)\{ \tau^{-1}, 1\}$ and  $\frac{2}{5}(2+\tau)\{ \tau^{-1}, 1\}$. The last two remaining cases in  Ref.~\cite{Wardman2011Thesis} are $\frac{8-\tau}{10}$ and $\frac{1+8\tau}{10}$, which are the linear combinations $\frac{1}{2}+\frac{3-\tau}{10}$ and $\frac{\tau}{2}+\frac{1+3\tau}{10}$, where the corresponding powers of $\tau+2$ are given by $1+3\tau=(\tau+2)\tau$ and $3-\tau=(\tau+2)\tau^{-2}$.

In conclusion, we note that  all  results  in Ref.~\cite{Wardman2011Thesis} and  Ref.~\cite{Keef:2009} have been rationalised here in a Coxeter group framework. For affine reflections perpendicular to  2-fold axes, the corresponding translations are inferred in $\mathbb{Z}[\tau]$; by contrast, for affine reflections perpendicular to 3- and 5-fold axes, an extra degree of freedom, $\gamma \in \mathbb{Q}$, occurs. However, we have shown here that a `natural' set of parameters occurs that, 
 once specified,  define a whole family of solutions and correspond to cases that occur in the context of applications in virology \cite{Keef:2009, Wardman2011Thesis}.


\subsection{Symmetrisability}

In the context of Kac-Moody algebras, it is often of interest to know if an asymmetric (generalised) Cartan matrix $A$ is  \emph{symmetrisable}. By this one means that there exist a diagonal matrix $D$ with positive integer entries and a symmetric matrix $S$ such that $A=DS$. 
This is of interest because in that case $S$ defines a scalar product on the simple roots, and much of the analysis carries through in the same way as for symmetric Cartan matrices. 


In order to generalise this notion of symmetrisability to our setting, we first note that the entries of the generalised Cartan matrices can be $\mathbb{Z}[\tau]$-integers (or even $\mathbb{Q}[\tau]$-rationals). Thus, the diagonal matrix may also take on positive values in $\mathbb{Z}[\tau]$ or  $\mathbb{Q}[\tau]$. 
In the Kac-Moody algebra context, the generalised Cartan matrix entries are integers, as they appear as powers of the adjoint action in the Chevalley-Serre relations. However, since we are working with a non-crystallographic root system for which no Kac-Moody algebras exist, we relax this condition here. 

Allowing for positive  $\mathbb{Q}[\tau]$-valued diagonal matrices, the families in Eq.~(\ref{CarH3affASfam2}),  Eq.~(\ref{CarH3affASfam3}) and  Eq.~(\ref{CarH3affASfam1}) considered earlier are indeed symmetrisable. The corresponding symmetric matrices are positive semi-definite (as are the Cartan matrices for Kac-Moody algebras of affine type) and  fulfil the determinant condition, which was to be expected. For instance, the diagonal matrices $\diag(\tau^2x^2, 1, 1, 1)$, $\diag(\frac{3}{4}\tau^2x^2, 1, 1, 1)$ and $\diag(\frac{1}{4}(\tau+2)x^2, 1, 1, 1)$ yield the following symmetrisations for Eq.~(\ref{CarH3affASfam2}),  Eq.~(\ref{CarH3affASfam3}) and  Eq.~(\ref{CarH3affASfam1}), respectively:
\begin{equation}
\begin{pmatrix}  2\tau^2\,{x}^{2}&0&x &0 \\ 0&2&-1&0 \\ x&-1&2&-\tau\\ 0&0&-\tau&2\end{pmatrix},\,\,\,\,   
 \begin{pmatrix} \frac{3}{2}\tau^2\,{x}^{2}&0&0 &x \\ 0&2&-1&0 \\ 0&-1&2&-\tau\\ x&0&-\tau&2 \end{pmatrix},\,\,\,\,   
 \begin{pmatrix}  \frac{\tau+2}{2}\,{x}^{2}&x&0 &0 \\ x&2&-1&0 \\ 0&-1&2&-\tau\\ 0&0&-\tau&2 \end{pmatrix}.  \label{CarH3affASfam2S}
\end{equation}

We note that in all three cases, there is only one choice for negative $x$ that makes the $(1,1)$-entry equal to $2$, i.e. which is such that the matrix can be interpreted in a Coxeter group setting as a   Cartan matrix,  and  the $(1,1)$-entry as a Coxeter exponent corresponding to a generator of a reflection. Note that this choice of $x$ also turns the diagonal matrix into the identity matrix, i.e. the original Cartan matrix is already symmetric. However, for this choice of $x$ the entries of the Cartan matrix are not necessarily $\mathbb{Z}[\tau]$-valued. In the first case,  we get $x=\sigma$,  and recover $H_3^{aff}$. For the second and third case, we obtain $x=\sqrt{\frac{4}{3}}\sigma$ and  $x=\sqrt{\frac{4}{5}(\tau-3)}$, respectively. Comparing with results in  Ref.~\cite{Keef:2009, Wardman2011Thesis}, we see that requiring symmetrisability in $\mathbb{Z}[\tau]$ excludes biologically important cases, in particular the Fibonacci series of affine extensions.

\section{Affine extensions of related non-crystallographic Coxeter groups}\label{sec_other}

We discuss here affine extensions of the non-crystallographic Coxeter groups $H_2$ and $H_4$, which are related to $H_3$ via the inclusion $H_2\subset H_3 \subset H_4$, and whose Cartan matrices are therefore structurally related.

\subsection{The case of $H_2$}\label{pointset}

A possible set of simple roots for $H_2$ is $\alpha_1=(1,0)$ and $\alpha_2=\frac{1}{2}(-\tau,\sqrt{3-\tau})$.
We consider affine reflections perpendicular to the highest root, or the bisector between the highest and an adjacent root. 
In the first case, the extended Cartan matrices are of the form:
\begin{equation}
A = \begin{pmatrix} 2&x&x \\ y&2&-\tau\\ y&-\tau&2 \end{pmatrix},\label{CarH2afffam1}
\end{equation}
which implies $xy=2-\tau=\sigma^2$. 
The solutions are again given by a Fibonacci family, which is the two-dimensional analogue to  the family  in Section \ref{sec_2fold}. 
The group generators of the affine extensions of $H_2$ and $H_3$ are  different, but since the affine extensions satisfy the same determinant constraint $xy=\sigma^2$, the two Fibonacci families are indexed by the same Fibonacci quadruplets.
Thus, $H_2^{aff}$ is the symmetric representative of this family, $x=y=\sigma$, and  a novel non-symmetric case is given, e.g. by $(x,y)=(\tau-2, -1)$.
As before, we obtain an exhaustive list of solutions via rescalings with $(\tau^k, \tau^{-k})$.

A second family is given by affine roots along the  bisector between the highest root and an adjacent root. Note that without loss of generality, we consider only one of these cases here as the geometry of the root system does not distinguish between the two bisectors due to the symmetry of the Cartan matrix  in $\alpha_1$ and $\alpha_2$. Hence:
\begin{equation}
A = \begin{pmatrix} 2&x&0 \\ y&2&-\tau\\ 0&-\tau&2 \end{pmatrix}, \label{CarH2afffam2}
\end{equation}
also with a  constraint on the product, $xy=3-\tau$, that we covered earlier in terms of the Fibonacci series for $H_3$. A solution is given by $(x,y)=(\tau-3, -1)$ and all others are again given by rescalings with  $(\tau^k, \tau^{-k})$.

\begin{figure}
	\begin{center}
       \begin{tabular}{@{}c@{ }c@{ }c@{ }}
		\begin{tikzpicture}
		\node (img) [inner sep=0pt,above right]
		{\includegraphics[width=3.5cm]{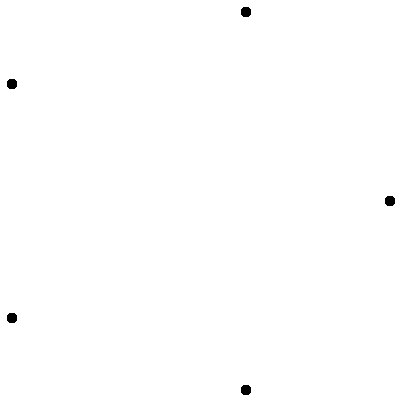}};
		\end{tikzpicture}&\hspace{1.5cm}
		\begin{tikzpicture}
		\node (img) [inner sep=0pt,above right]
		{\includegraphics[width=3.5cm]{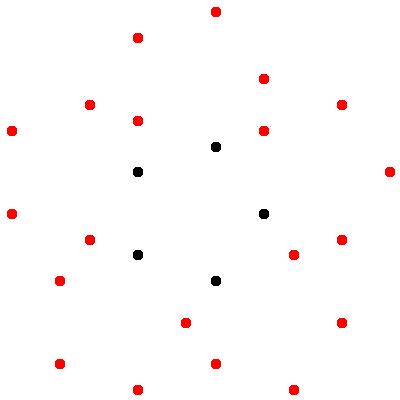}};
		\end{tikzpicture} &\hspace{1.5cm}

		\begin{tikzpicture}
		\node (img) [inner sep=0pt,above right]
		{\includegraphics[width=3.5cm]{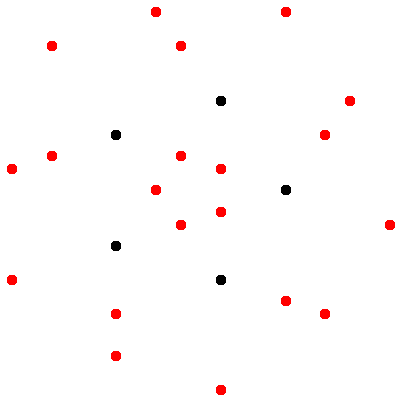}};
		\end{tikzpicture}
	\vspace{0.5cm}
	\\
	(a)&(b)&(c)	\vspace{0.5cm}\\

			\begin{tikzpicture}
			\node (img) [inner sep=0pt,above right]
			{\includegraphics[width=3.5cm]{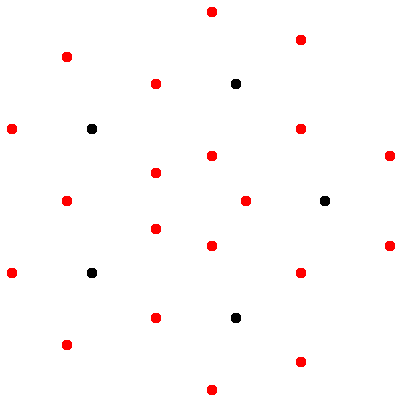}};
			\end{tikzpicture}&\hspace{1.5cm}
			\begin{tikzpicture}
			\node (img) [inner sep=0pt,above right]
			{\includegraphics[width=3.5cm]{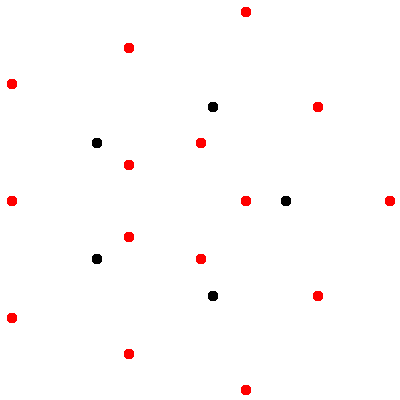}};
			\end{tikzpicture} &\hspace{1.5cm}

			\begin{tikzpicture}
			\node (img) [inner sep=0pt,above right]
			{\includegraphics[width=3.5cm]{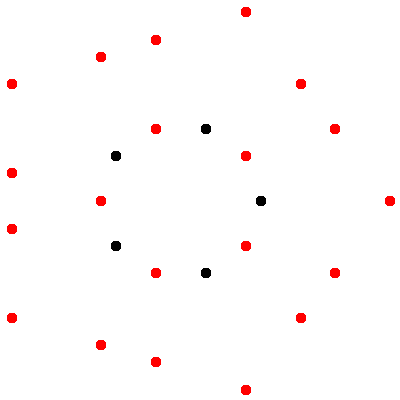}};
			\end{tikzpicture}
	\vspace{0.25cm}
		\\
	(d)&(e)&(f)\\
 	\vspace{0.25cm}
  \end{tabular}
  \caption[Hi]{Visualisation of the action of the affine extended groups on a pentagon (a): panels (b) and (c) depict two translations along a bisector with translation lengths $\sqrt{2+\tau}$ and $\sqrt{3-\tau}$, respectively. (d-f) shows translations along the highest root of length $-\sigma$, $1$ and $\tau$, respectively. The cardinality of the point sets is 25, except for (e), which has cardinality 20 and corresponds to $H_2^{aff}$.  }
\label{figbis}
\end{center}
\end{figure}

We note that in both cases the determinant constraints are integer-valued, and hence the solutions are purely given in terms of the Fibonacci and Lucas integers. The identity  Eq.~(\ref{angle}) determines the angle between the affine root $\alpha_0$ and each of the simple roots $\alpha_1$ and $\alpha_2$ if $xy=2-\tau$ as $2\pi/5$, as expected from Ref.~\cite{Twarock:2002a}, and  as $3\pi/10$ if  $xy=3-\tau$, as  expected for  the bisectors of the roots.

For the `symmetric' series in Eq.~(\ref{CarH2afffam1}), the  lengths are $\tau^k$, and for the `asymmetric' series  Eq.~(\ref{CarH2afffam2}), the  lengths are $\sqrt{3-\tau}\tau^k$.  
We visualise these affine extended groups via their action on a pentagonal configuration in Fig. \ref{figbis}. Fig. \ref{figbis} (b) and (c) show two translations from our classification of length $\sqrt{2+\tau}$ and $\sqrt{3-\tau}$, respectively,  along a bisector, and Fig. \ref{figbis} (d-f) show translations parallel to the highest root of length $-\sigma$, $1$ and $\tau$, respectively. Note that in all cases, the cardinality of the point array is smaller than that for a random translation, which corresponds to 30, rather than 25 as in (b-d) and (f) or 20 as in (e).

\subsection{The case of $H_4$}

For completeness, we also briefly discuss  extensions 
for $H_4$. The already familiar case of $xy=\sigma^2=2-\tau$ occurs for extensions along the highest root $\alpha_H$
\begin{equation}
A_1 = \begin{pmatrix} 2&x&0&0&0 \\ y&2&-1&0&0 \\ 0&-1&2&-1&0 \\ 0&0&-1&2&-\tau \\ 0&0&0&-\tau&2 \end{pmatrix},\label{CarH4affASfam1}
\end{equation}
where the index on $A$ denotes the column of the non-zero entry $x$ in the extended Cartan matrix. 
The symmetric case $x=y$ corresponds to $H_4^{aff}$, and all other solutions can be derived
from $(x,y)=(\tau-2, -1)$ via rescaling by $(\tau^k, \tau^{-k})$. 

 Moreover, the families
\begin{equation}
A_3 = \begin{pmatrix} 2&0&0&x&0 \\ 0&2&-1&0&0 \\ 0&-1&2&-1&0 \\ y&0&-1&2&-\tau \\ 0&0&0&-\tau&2 \end{pmatrix}  \text{ and } \,\,\,\,\,A_4 = \begin{pmatrix} 2&0&0&0&x \\ 0&2&-1&0&0 \\ 0&-1&2&-1&0 \\ 0&0&-1&2&-\tau \\ y&0&0&-\tau&2 \end{pmatrix}\label{CarH4affASfam3}
\end{equation}
have the determinant constraints $xy=\frac{1}{3}(5-3\tau)$
and $xy=\frac{1}{2}(5-3\tau)$, respectively. Since {$5-3\tau=(2-\tau)^2=\sigma^4$}, there is a symmetric  solution $(-2,1; -2,1)$, from which we can generate all solutions via    $(\tau^k, \tau^{-k})$ rescalings.

Finally,
\begin{equation}
A_2 = \begin{pmatrix} 2&0&x&0&0 \\ 0&2&-1&0&0 \\ y&-1&2&-1&0 \\ 0&0&-1&2&-\tau \\ 0&0&0&-\tau&2 \end{pmatrix}\label{CarH4affASfam2}
\end{equation}
has determinant constraint  $xy=\frac{1}{5}(7-4\tau)=\frac{1}{5}(2-\tau)(3-\tau)$. In this case, all solutions can be derived from   $(-1,0; -7,4)$ and $(-7,4; -1,0)$  in terms  of Fibonacci and Lucas numbers.

\section{Conclusions}\label{concl}

We have extended the Kac-Moody formalism for affine extensions of root systems to the case of asymmetric $\mathbb{Z}[\tau]$-valued affine Cartan matrices. 
A consistency condition on the new root has been presented, and it has been shown how the affine reflection associated with the additional root gives rise to translations and twist translations along the direction of the affine root. The case of $H_3$ has been discussed in detail, as it is the most relevant for practical applications in biology and carbon chemistry. We have considered extensions along the icosahedral symmetry axes, and classified the allowed translations in terms of a   Fibonacci recursion relation. Thus, we have rationalised results in Refs ~\cite{Keef:2009, Wardman2011Thesis} at the Coxeter group level. 

Finally, we have discussed similar affine extensions for $H_4$, and for $H_2$. For the latter, we have visualised  the action of the  extended groups geometrically in a vector space representation in terms of point arrays. In this framework,  one can consider linear combinations of translations; in particular, we note that translations along 3-fold and 5-fold axes of icosahedral symmetry can be obtained as linear combinations of translations along  three (respectively five) root vectors, and analogously, one could  also contemplate more general linear combinations of root vectors. 

{We conclude that a wide range of empirical observations in virology can be explained by  affine Coxeter groups. 
For example, 3D point arrays analogous to those discussed in Section ~\ref{pointset} predict the architecture of viruses and fullerenes. We have developed here a group theoretical framework that allows for these structures to be modeled in terms of (asymmetric) affine extensions of non-crystallographic Coxeter groups. }


\section*{Acknowledgements}
We would like to thank Prof Vladimir Dobrev for a careful reading of the manuscript and many valuable suggestions. RT gratefully acknowledges support via a Research Leadership Award from the Leverhulme Trust that has provided funding for PPD.


\end{document}